\documentclass[aip,amssymb]{revtex4-1}
\usepackage{graphicx}
\usepackage{xcolor}
\draft 

\begin{document}


\title{Piezoelectric tiles for passive flow rate monitoring across a surface} 



\author{S. Hales Swift}
\author{Ihab F. El-Kady}
\affiliation{Sandia National Laboratories, Photonic and Phononic Microsystems}


\date{\today}

\begin{abstract}
We introduce a method for measuring the velocity of turbulent fluid flow passing through a pipe using piezoelectric tiles without penetrating the pipe, and without having previously designed the pipe to easily allow monitoring. To measure the flow, the vibrations induced on the pipe by the fluctuating pressure loading induced by the turbulent flow are measured and compared across flow speeds to establish effective invertible relationships from vibration to velocity. Measurements are reported for instrumented pipes transporting, in separate experiments, water and air. The water experiment was able to resolve linear velocity differences on the order of 1~cm/second, while the air experiment was able to resolve on the order of 15~cm/sec. Turned inside out, a similar system might be used to assess \emph{external} flow velocity, determine differential velocities on opposite sides of a body traveling through air and water, and thus provide navigational data in the form of speed and attitude/angle of attack information. Although this approach is prototyped for a single sensor, it is likely to benefit substantially from the noise suppression possible when employing an array of sensors.
\end{abstract}


\maketitle 

\section{Flow detection/speed monitoring}
The ability to externally characterize flow passing through a pipe has a number of potential applications including leak or obstruction detection~\cite{Korlapati:2022}, industrial process control in which chemicals must be kept in appropriate ratios, or monitoring HVAC flow rates through a building or fuel flow through a system. In some cases, the ability to measure flows without penetrating the pipe itself may be particularly important such as the case when monitoring the flow rate of chemicals that might harm sensors exposed to them, or where the sensor itself might contaminate materials requiring high purity, or where penetrations may lead to an unacceptable risk of leak development, or perturbation of the flow itself. While effective methods exist for pipes where such monitoring is an intended feature of the system~\cite{Ren:2022}, and for systems involving excitation~\cite{Kim:1996} for the purpose of this effort it is not assumed that such access must have been planned in advance, or that excitation is possible: one might, for example, decide that monitoring access is needed for a legacy or established system. We will demonstrate a method for adding piezoelectric-based vibration monitoring instrumentation to a pipe that was not specifically prepared to accept such instrumentation, and using the vibration instrumentation to monitor the fluid flow rate through the pipe. The central premise of this work is that provided a piping system is transporting a fluid with effectively constant material properties, the character of the turbulence-induced vibration in the pipe can be inverted to obtain a close estimate of the velocity. A similar approach was taken by Evans~\cite{Evans:2004} with accelerometers; however, we show results for raw piezoelectric tiles, resulting in both some additional processing needs and associated precautions.

We further note by way of motivation that a system that can monitor the flow through a closed pipe can be turned inside out to instead monitor the exterior flow over a closed body passing through air or water. The ability to monitor exterior flow rates thus provides a \emph{de facto} speedometer to air or water vehicles, including in harsh environments where maintaining a sealed exterior is important for sustained vehicle function, e.g., for undersea vehicles, or for vehicles traveling at high altitudes or speeds. The use of piezoelectric tiles contributes minimal additional mass and volume in keeping with the size, weight, and power constraints of most practical application. Extended still further, the exterior version of this concept can be used to monitor differential flow rates over opposing portions of the vehicle providing attitude or angle of attack information, and can thus serve as an aid to an inertial navigation system (INS) providing important correctives for the navigation process. Left to their own devices, INSs begin to accumulate errors when real-position navigational updates become unavailable~\cite{Brody:2023}, whether through malfunction, RF interference, poor undersea transmission, positioning with poor celestial sight lines, or other circumstances that interfere with receiving unimpeded global navigation satellite system signals. Flow velocity estimation using vibration sensors on the side of a plate not exposed to the flow has been explored~\cite{Swift:2025}, as have attitude and velocity assessment using internally carried sensors for high speed (hypersonic) flow, and appears to be quite accurate~\cite{Smith:2026}. The present paper, however, will extend these results in the opposite direction to lower speed scenarios, and to the additional case where water rather than air is the fluid of interest.

It is in this context that we consider the turbulent flow through a closed pipe. Turbulence in a pipe flow occurs when the Reynolds number
\begin{equation}
    Re=\frac{\rho v D}{\mu}=\frac{v D}{\nu}>4000,
\end{equation}
where $D$ is pipe diameter, $\rho$ is density, $v$ is velocity, $\mu$ is viscosity and $\nu$ is kinematic viscosity; for water at $20^o$C, $\nu_{water}\approx1\times10^{-6}~\mathrm{m^2/s}$, and for air at $20^o$C, $\nu_{air}=1.506\times10^{-5}~\mathrm{m^2/s}$ so the equation amounts to a prediction of turbulent flow when the condition in Table~\ref{tab:turbulence_criteria} are satisfied. These criteria are also expressed in terms of volume velocities, $V=v\pi(D/2)^2$, for circular pipes.

\begin{table}[h]
    \centering
    \begin{tabular}{|c|c|c|}
    \hline
    Substance & Turbulence criterion&Volume velocity criterion\\
    \hline
        Water ($20^o$C)& $vD>0.004~\mathrm{m^2/s}$ & $V>0.00314D$~m$^3$/s\\
        Air ($20^o$C)& $vD>0.06024~\mathrm{m^2/s}$ & $V>0.0473D$~m$^3$/s\\
    \hline
    \end{tabular}
    \caption{Criteria for expecting turbulent flow in pipes for air and water.}
    \label{tab:turbulence_criteria}
\end{table}
In practice, turbulence can be further encouraged by roughening walls, providing trips, and so forth if one has access to the interior of the pipe through which the water is flowing. In this work, we assume no such access will be available in applying this approach, and so have not made these a main focus. In general, however, a smaller pipe better encourages turbulence. On the other hand, piezo transducers are easiest to properly affix to a wide diameter pipe on purely geometric grounds because a wider pipe is less prone to curve away from the (flat) transducer. We instrumented pipes with a nominally 8-cm internal diameter, which would mean that turbulent flow conditions with water or air would require a pump or fan, respectively, capable of producing the values given in Table~\ref{tab:turbulent_volume_velocities} though, in practice, an ability to vary the flow in an adequate range beyond these minimum values is preferred.
\begin{table}[h]
    \centering
    \begin{tabular}{|c|}
    \hline
    Volume velocity requirement\\
    \hline
        $V_{water}>0.0002512~\mathrm{m^3/s}$ \\
        $V_{air}>0.003784~\mathrm{m^3/s}$ \\
    \hline
    \end{tabular}
    \caption{Minimum volume velocity values for turbulent flow of water or air at $20^o$C through an 8-cm-internal-diameter pipe.}
    \label{tab:turbulent_volume_velocities}
\end{table}

\section{Apparatus}
For both air and water, piezoelectric tiles were affixed using 3M Scotch Weld DP460 Epoxy to an 8-cm-inner diameter acrylic pipe. Resistors were used to short the tiles to prevent the build up of excess voltage (a known issue given the capacitance of piezoelectric tiles). Data acquisition was performed using an National Instruments cDAQ chassis with 9220 and 9215 acquisition cards. Sampling for these experiments was at $f_{samp}=25,000$~Hz.

\begin{figure}
    \centering
    \begin{tabular}{cc}
         \includegraphics[width=0.48\linewidth,clip=true,trim=0mm 30mm 0mm 0mm]{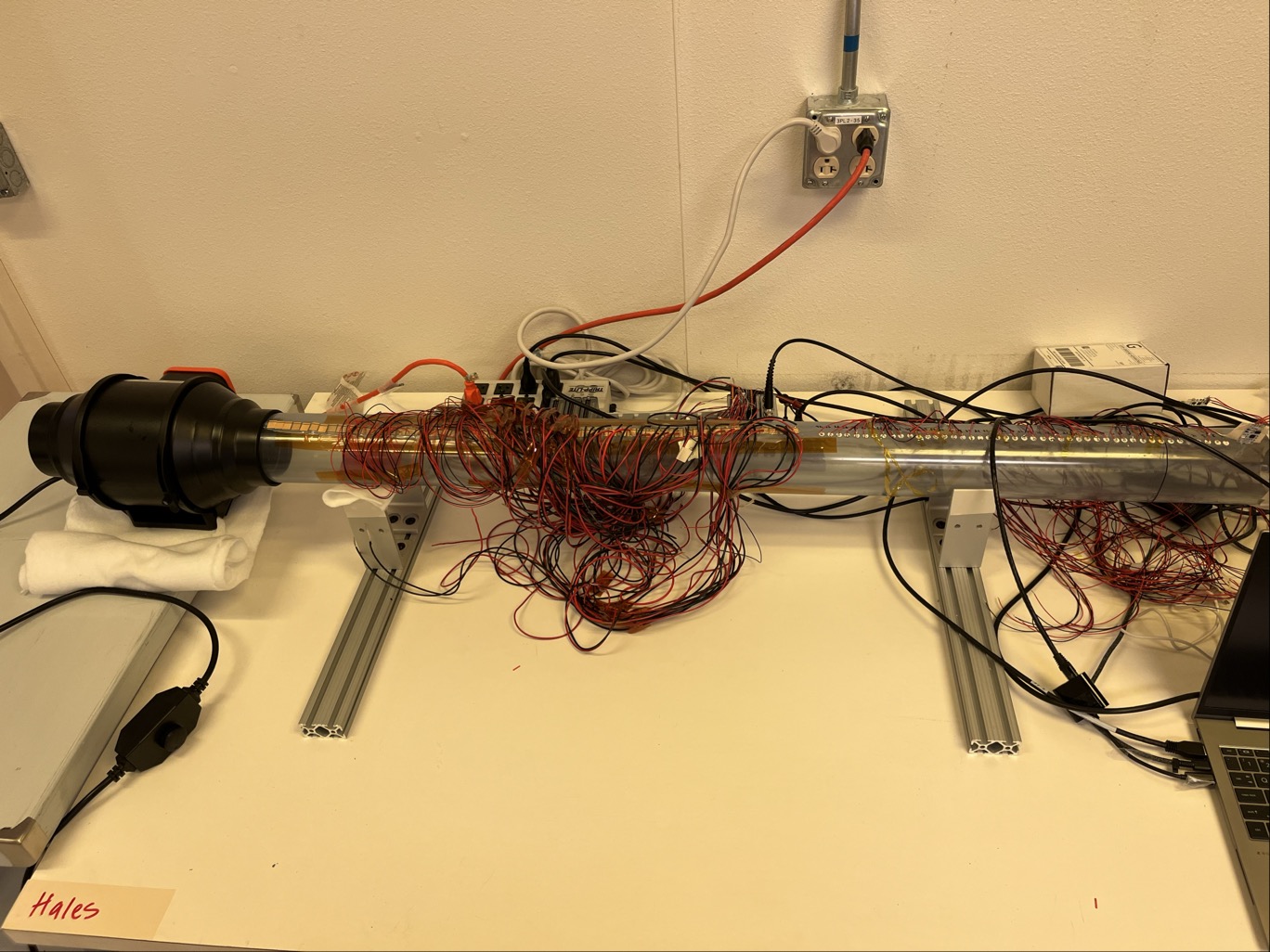}&\includegraphics[width=0.48\linewidth,clip=true,trim=0mm 50mm 0mm 4mm]{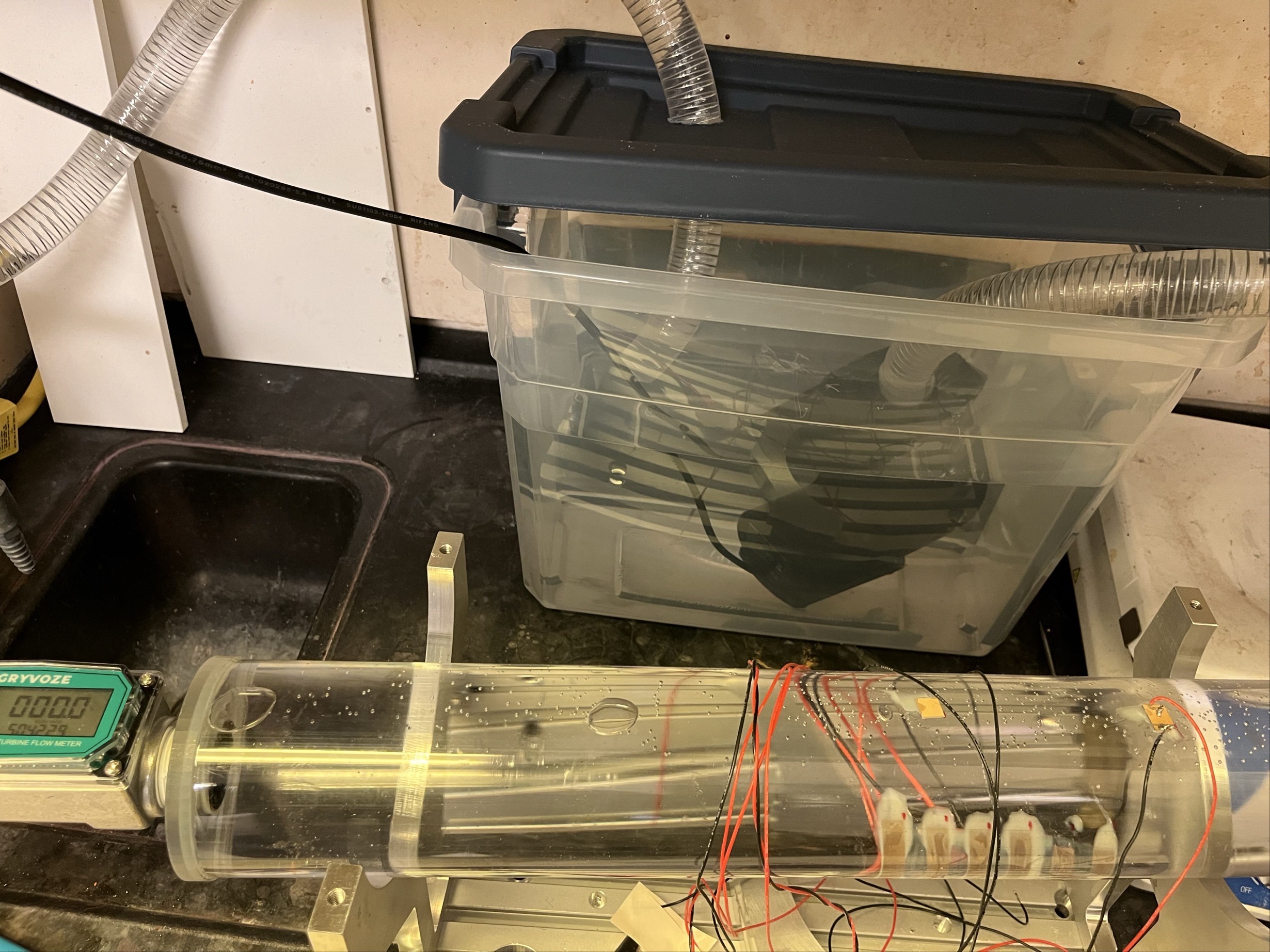}
    \end{tabular}
    \caption{(Left) The air tube setup. (Right) The water tube setup.}
    \label{fig:tubes}
\end{figure}

For the water tube, end caps were cut out of acrylic sheet material using a Roland MDX-540 CNC machine, with finish machining performed using a Dremel tool, and other hand tools. The end caps were fixed in place using DP460 epoxy. A 1-inch PVC steel wire reinforced hose connected the pipe to a water pump in a nearby reservoir. The same hose material provided the return feed to the reservoir, and this was suspended to avoid inducing turbulent air-water surface mixing that was a suspected source of noise in early experiments. An HG181-34W brushless aquarium pump provided flow velocities between 30.4~L/min and 40.3~L/min as measured by a Gryvoze 1 inch digital turbine fuel flow meter, as shown in Fig.~\ref{fig:tubes} (Right). Threads were sealed with PTFE thread seal tape to avoid leaks.

For the air tube, air flow was provided by a VT-FL14A ventilator fan, which was operated with the suction end facing the tube so that flow disturbances linked to the blade passage frequency would not provide a constant source of periodic excitations in competition with the turbulence that was the intended target of measurement as seen in Fig.~\ref{fig:tubes} (Left). After recognizing the role of mechanical conduction of vibration, the fan was kept on a separate adjacent table, and positioned so that there was a small air gap to prevent direct transmission of vibration between the fan and the tube. Both the fan and the tube were placed atop small absorptive material stacks to further discourage transmission via conductive mechanical vibration paths. A Modern Device Wind Sensor Rev. P6 analog-output hot wire anemometer was used with a conventional 12-V power source to provide flow estimates for the air tube setup. A small slit was drilled through the top of the tube through which to situate the anemometer within the flow. The use of an anemometer with an analog output proved highly useful because it made it possible to record simultaneous data for both the anemometer and the piezoelectric tile to facilitate meaningful analysis of the relationship between the two variables.

\section{Water Tube}
\subsection{Initial efforts}
We obtained flow volume velocity estimates from the flow meter for the different pump settings, and these are given in Table~\ref{tab:volume_velocities}. 30.4~L/min is equivalent to 0.00051~m/s, which is approximately twice the flow we need in order to obtain turbulent conditions. Consequently, turbulent flow is expected for all pump settings. The velocity varies across pump settings from a baseline value to a value roughly 33\% greater. We were not able to measure fine variations in flow velocity using the flow meter we used as these changes were smaller than the gauge's measurable differences. Although the measurements using the flow meter were informal---we simply watched the output of the gauge for a period of time at each pump setting and observed and recorded that it adopted a consistent value---they turned out to be highly consistent with variation from the stated values uncommon. Unfortunately, the gauge was not able to distinguish between the two highest pump settings.
\begin{table}[h]
    \centering
    \begin{tabular}{c|cccccccccc}
        \hline
         Setting&1&2&3&4&5&6&7&8&9&10\\
         Volume Velocity (L/min)&30.4&32.0&32.9&34.5&35.3&36.1&37.8&39.4&40.3&40.3 \\
         Velocity (m/s)&0.1008&0.1061&0.1091&0.1144&0.1170&0.1197&0.1253&0.1306&0.1336&0.1336\\
         \hline
    \end{tabular}
    \caption{The volume velocities for the several pump settings of the hygger$R$ brushless aquarium pump HG181-34W as read out using the Gryvoze digital turbine flow meter, and their associated linear velocities.}
    \label{tab:volume_velocities}
\end{table}

Preliminary measurements of vibration of the tube at each pump setting suggested that power line electromagnetic interference (EMI) contamination was present. Accordingly, several mitigation solutions were evaluated, including the use of a LMS filter with an antenna reference channel for active noise control; however, the solution that was ultimately employed was removal of EMI harmonics (evaluated with the pump off) in the frequency domain.

\begin{figure}
    \centering
    \includegraphics[width=0.75\linewidth]{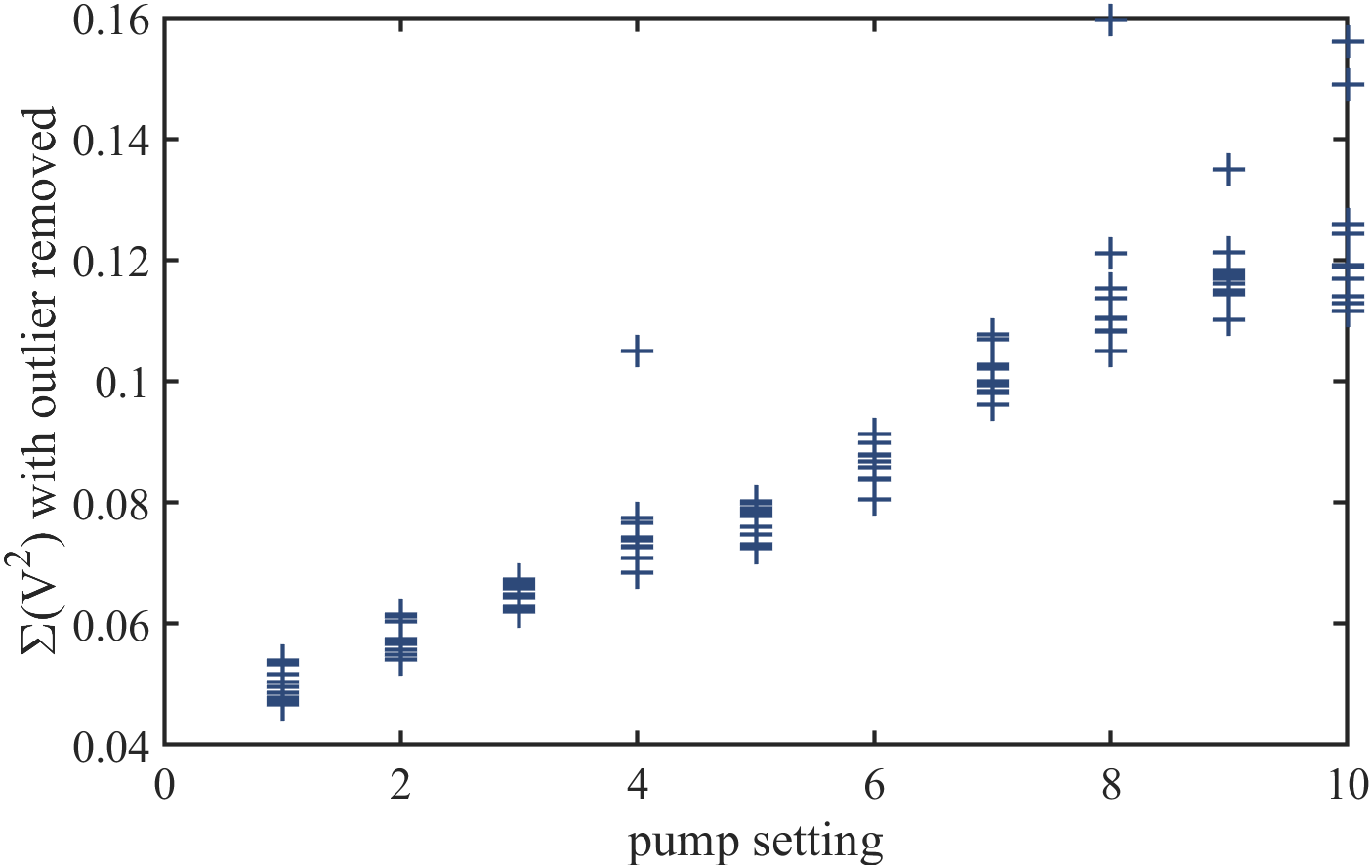}
    \caption{Processed measured vibration power for individual trials at each pump setting.}
    \label{fig:vibration_settings1}
\end{figure}

A preliminary measurement was conducted across all conditions with the pump setting ascending and then descending to evaluate consistency of performance in the pump (LMS filtering was used for EMI control at this stage). The results for ascending and descending cases are shown in Fig.~\ref{fig:vibration_settings1}, where it can be seen that most of the vibration response (in terms of voltage squared, $V^2$) measurements group tightly with one another based on their common pump settings, such that an uncontaminated signal of adequate length would likely be sufficient with appropriate processing to determine the pump setting with high confidence in the absence of other confounding conditions.

\begin{figure}[h]
    \centering
    \includegraphics[width=0.75\linewidth]{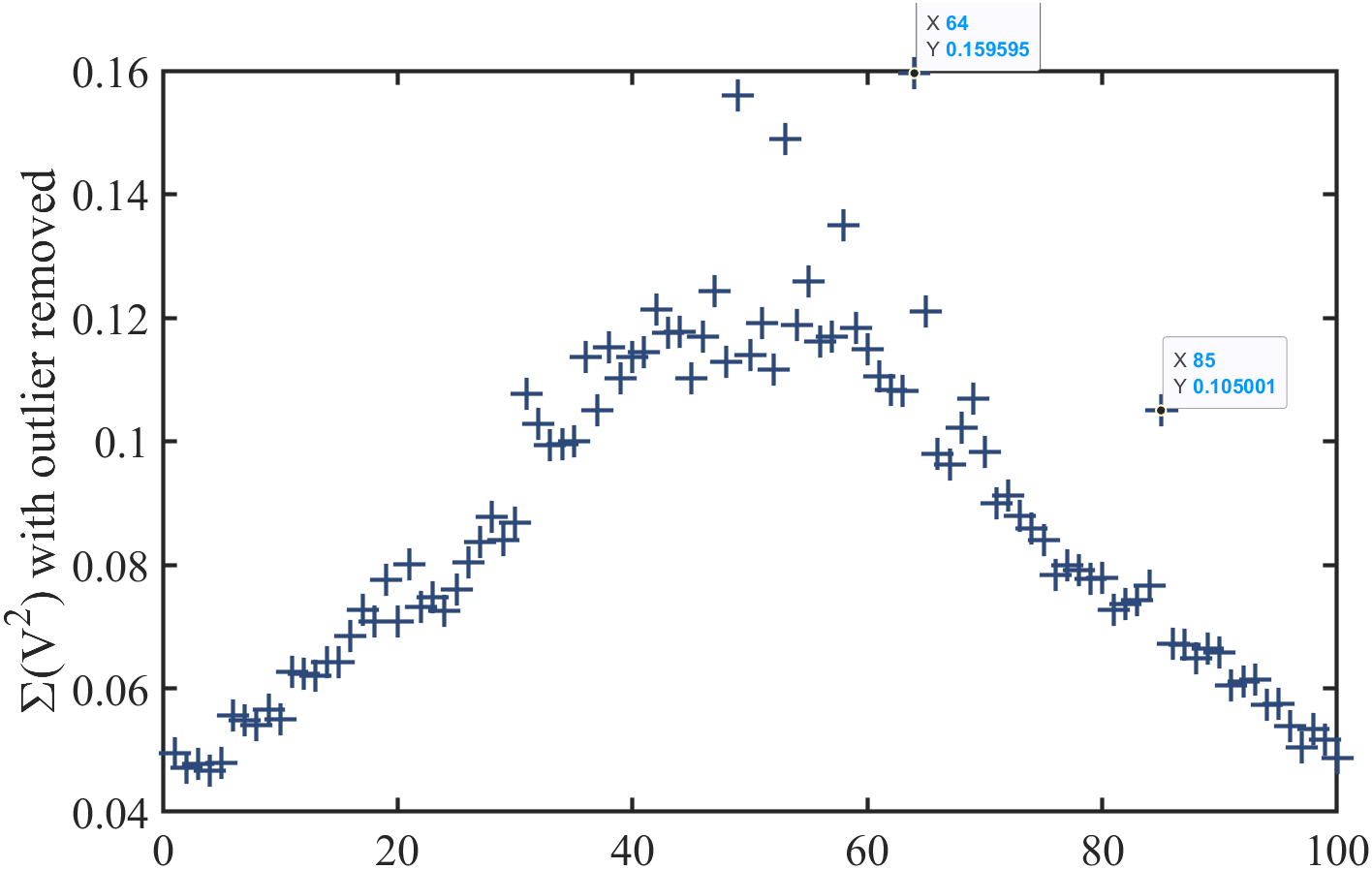}
    \caption{The vibration response of the tube is measured for ascending and descending pump settings.}
    \label{fig:ascend_descend1}
\end{figure}

\begin{figure}[h]
    \centering
    \begin{tabular}{cc}
         \includegraphics[width=0.75\linewidth]{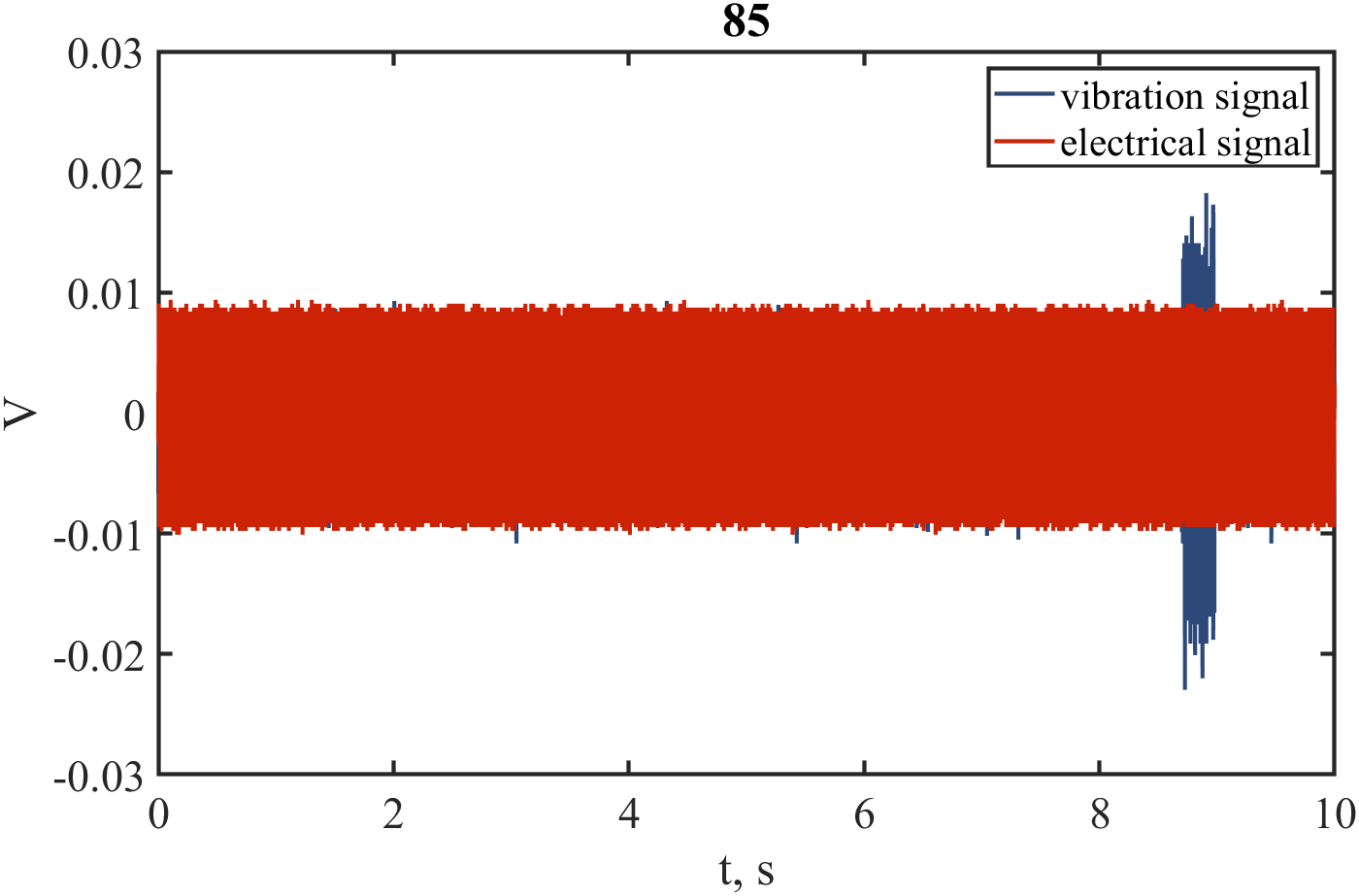}\\
    \end{tabular}
    \caption{Time series for the electrical signals and vibration signals associated with an outlier.}
    \label{fig:temporal_outliers}
\end{figure}

However, several clear outliers were present, two examples of which are marked in Fig.~\ref{fig:ascend_descend1}.
Looking at the time series for one the marked points (the other is similar in character), it appears clear that there are abnormal temporal events in both cases as shown in Fig.~\ref{fig:temporal_outliers}; because we have both the electrical and vibration time series, we know that the abnormality is not due to electromagnetic interference, but appears to be due to some sort of transient vibration phenomenon. A delay was implemented between changing the pump condition and initiating the measurement, so we can likely also rule out transient events associated with transitions between pump conditions. This leaves the possibility of other sharp random vibration events in the local environment, which is a working lab. Because these events tend to be relatively brief in time, and the level appears to be stable otherwise, a temporal outlier rejection procedure appeared to be a promising approach to resolving the normal vibration behavior while ignoring unusual transients. 

\subsection{Temporal outlier removal}
Examination of the squared voltage smoothed over $0.2\bar{6}$ seconds, $V^2_{\tau = 0.2\bar{6}}$, revealed that the outliers manifested as isolated positive deviation in squared voltage, while most of the signal was unaffected; consequently, the outlier cases exhibited significant skewness in $V^2_{\tau = 0.2\bar{6}}$. Consequently, a criterion was implemented such that samples with $\mathrm{abs}(\mathrm{Sk}\{V^2_{\tau = 0.2\bar{6}}\})>0.5$ were rejected, with sampling continuing until five valid samples at each setting were acquired in ascending and descending order (to address any possible hysteresis) for a total of ten samples at each pump setting.

\begin{figure}[h]
    \centering
    \includegraphics[width=0.75\linewidth]{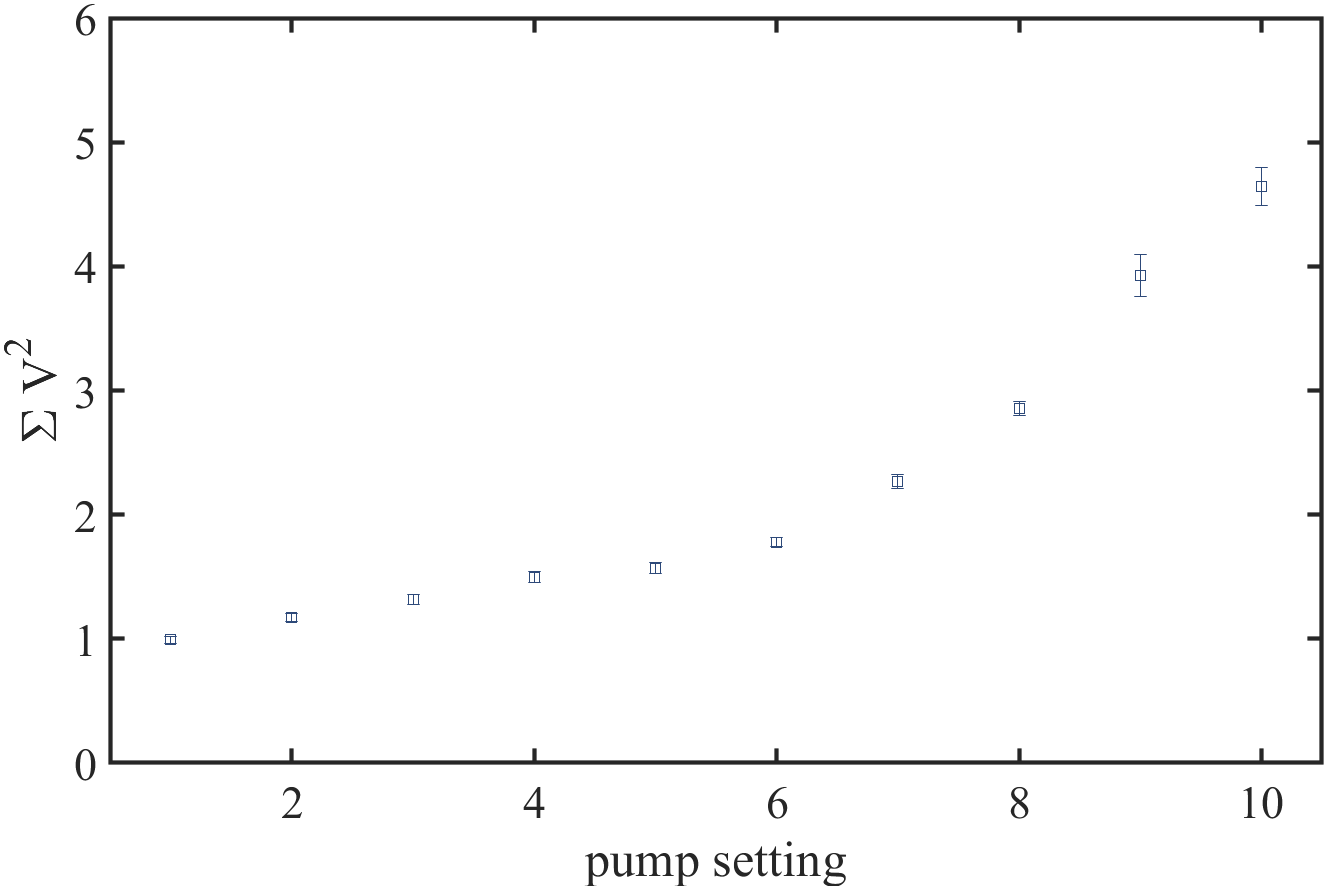}
    \caption{After applying EMI filtering and the skewness-based outlier rejection techniques, the variability in the samples associated with each pump setting (standard deviation are shown) is modest, and allows inversion with a high degree of confidence.}
    \label{fig:vibration_setting}
\end{figure}

After obtaining a set of low-skewness samples following the procedure of the previous paragraph, the outliers were effectively removed resulting in the narrow standard deviation seen in Fig.~\ref{fig:vibration_setting}, where both skewness rejection sampling and EMI filtering were successfully used to show relatively distinct levels of vibration associated with each setting, and associated flow velocity. 
Notably, the processed mean squared vibration values associated with the different settings are different enough that their standard deviations bars do not overlap except for settings 4 and 5. This suggests that given an adequate uncontaminated sample of the vibration it should be possible to invert for pump setting based on the consistency of turbulence response at each of the set output volume velocities. The vibration measurement was also able to resolve differences between settings 9 and 10, which the gauge was not able to resolve; we suspect that this reflects a limitation in the flow gauge in distinguishing this particular increment of flow rate. Because the flow gauge had limited ability to resolve flow rates, we were not able to evaluate the the exact accuracy of this approach in water beyond the volume velocity categories and associated linear velocity values that were typical (and virtually constant from the gauge's perspective) for the different pump settings. However, from a linear velocity perspective, these are differences of less than a centimeter per second. Consequently, the ability of this approach to finely resolve flow velocities appears to be excellent provided uncontaminated samples can be assured. Further efforts should focus on the ability to generalize this method to the external flow case given the differences between internal and external flow over a cylindrical surface.


\section{Air Results}
Preliminary analysis with a handheld anemometer (Vantech$\circledR$ VA103) revealed that individual fan settings were not consistently repeatable, did not lead to entirely stable air flow rates, and in fact showed some hysteresis effects, so these could not be used effectively as velocity classes. The role of conductive transmission of vibration was realized iteratively as different changes were made ending with the fan uncoupled and isolated from the tube on a separate table as described previously. Additionally, the difficulty of time aligning readings from a handheld anemometer with vibration results were substantial and pointed to the need to simultaneously record the velocity and vibration data in order to enable meaningful investigation of the relationship between the two variables

\begin{figure}[h]
    \centering
    \includegraphics[width=0.75\linewidth]{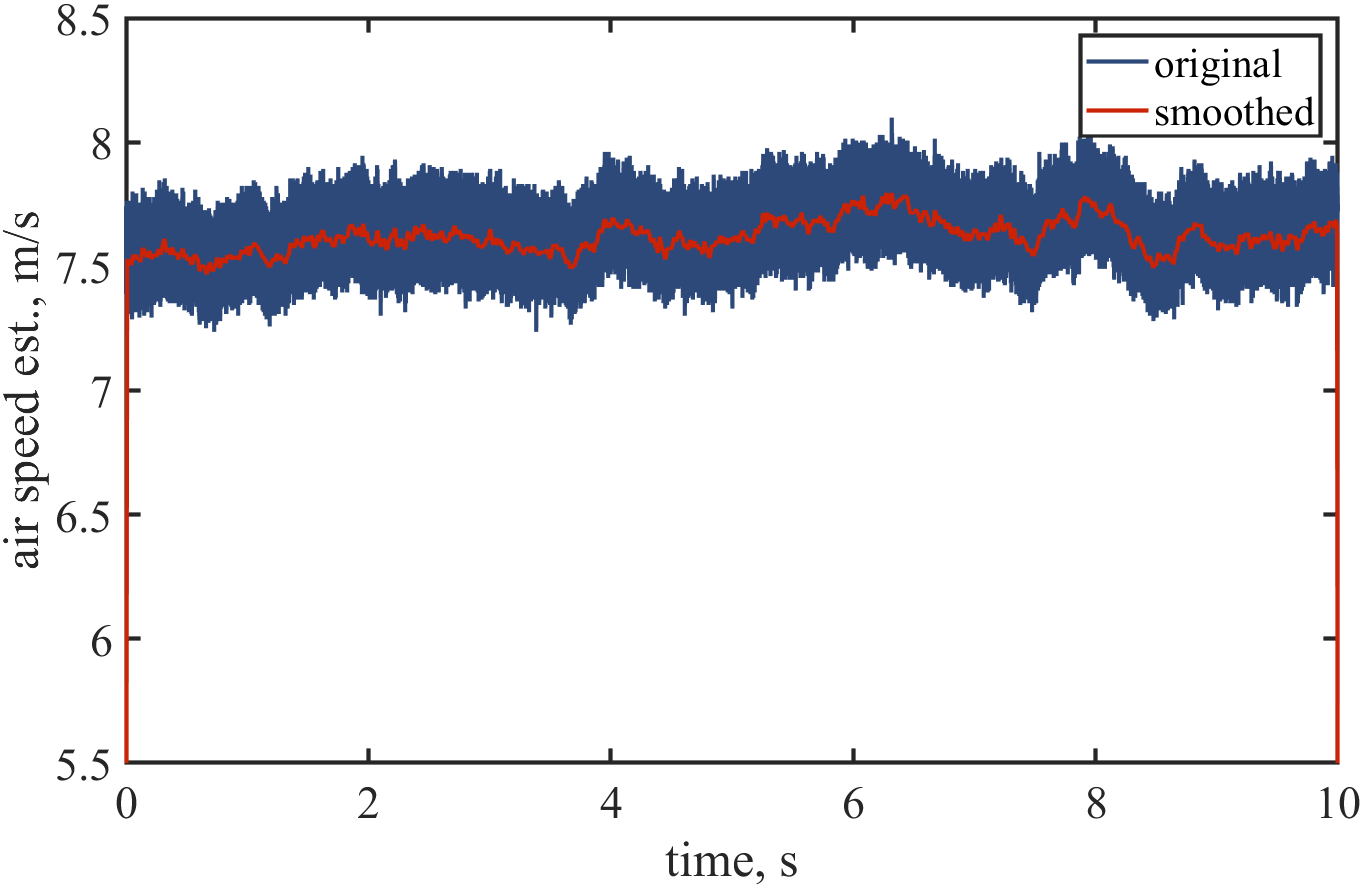}
    \caption{With the fan at a constant setting, the airflow measured by the anemometer continues to vary.}
    \label{fig:fan_variability}
\end{figure}

The analog anemometer was subject to some electromagnetic interference from powerline noise; to minimize this, we used a moving average over $1/60^{th}$ of a second to naturally filter against these and other higher-frequency interference sources. Even when operating at a constant setting, the fan displayed noticeable variability as shown in Fig.~\ref{fig:fan_variability}, where both the raw and smoothed velocity estimates are shown on the same plot.


Although the piezoelectric tiles were shorted with resistors to limit non-zero mean behavior, significant excursions existed from this ideal. Consequently, a 1/3-second Hann window normalized by its sum was used in a convolution with the voltage time history to track local non-zero mean behavior and remove it from the signal. The ends where the convolution did not have fully whetted data were extrapolated from the nearest points where complete data was available, and this was subtracted from the signal to compensate for deviations from locally zero-mean behavior. Suspected electromagnetic interference was then removed from the signal in the frequency domain based on a preliminary analysis of the spectrum under ambient (fan off) conditions, resulting in a decrease in RMS level of 11.3\% in the case we measured; consequently, this step seems fairly important in mitigating EMI.

\begin{figure}
    \centering
    \includegraphics[width=0.75\linewidth]{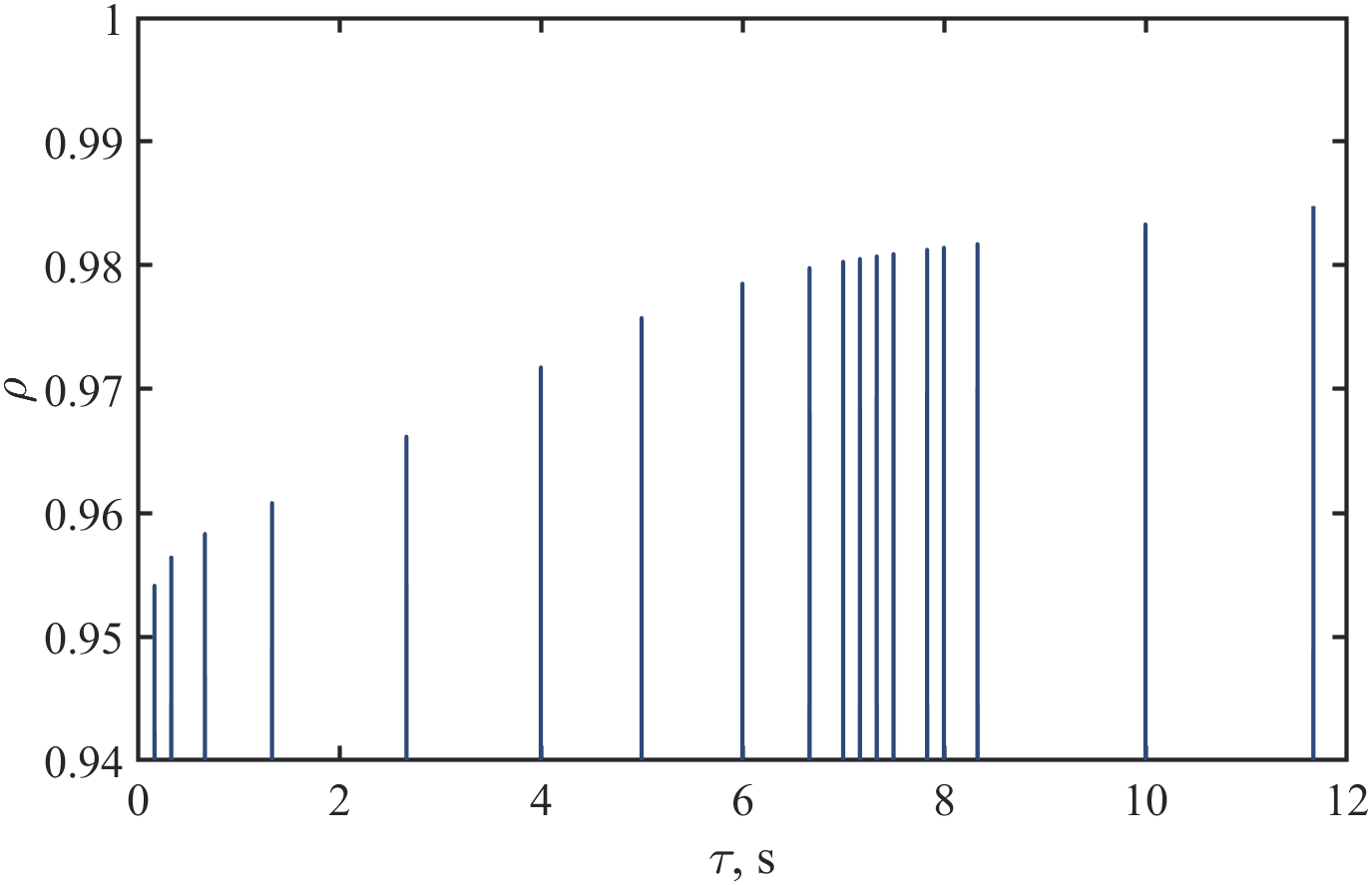}
    \caption{Correlation for different amounts of time averaging. Over the intervals considered, correlation increased with length of time averaging despite.}
    \label{fig:rho_tau}
\end{figure}

In Fig.~\ref{fig:rho_tau}, the average length and the resultant correlation coefficient between the velocity and the processed vibration power measurement time series are shown together. Longer time averages appear to improve the quality of the relationship between the velocity and the vibration power. Additionally, transient airflow events are a less frequent part of the operation of most systems. Consequently, we will look at longer time averaging periods of 10 seconds. This precise length is somewhat arbitrary, but represented a compromise between the desire for strong predictive ability from the vibration level to the velocity on the one hand, and the need for adequate time record lengths to support those durations, and time specificity on the other.

For our chosen averaging period of $\tau=10$~seconds, delaying the signal from the anemometer by 0.1865~seconds yielded the highest correlation. The data this was based on involved faster fluctuations in fan setting in the hopes of ensuring that the method ultimately settled upon represented adequate tracking between the two variables.

\begin{figure}
    \centering
    \includegraphics[width=0.75\linewidth]{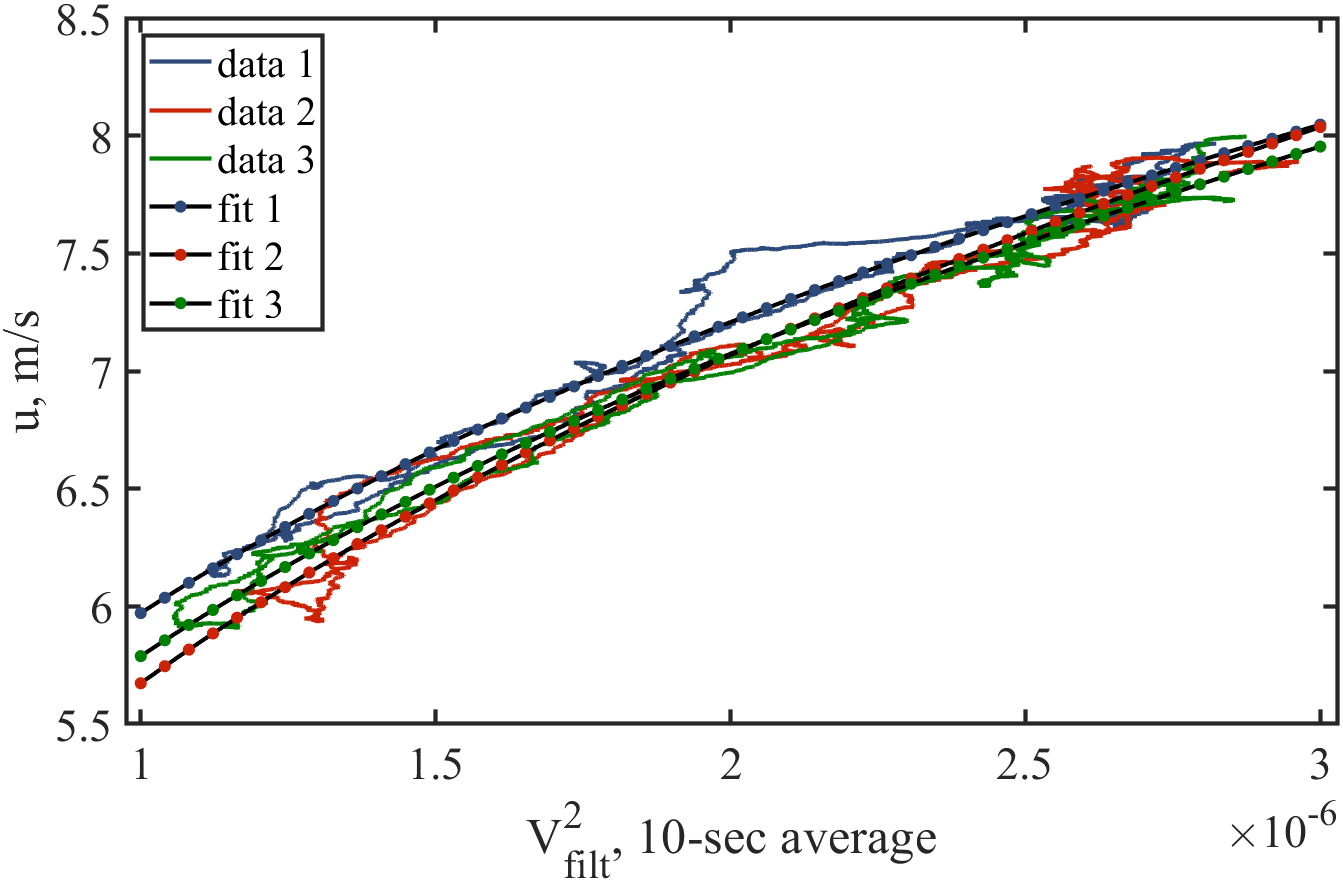}
    \caption{Processed and smoothed data from the air tube relating the anemometer-based velocity and the squared vibration signal and associated exponential fits to the processed data.}
    \label{fig:fits}
\end{figure}

Having settled the processing scheme to be used, we took a new data set using slower variation in order to make sure that we hit and settled on the maximum and minimum function settings of the fan and arrived at and departed from those settings slowly enough to ensure that we got representative data that could be appropriately averaged over longer periods of time. These data sets are accordingly longer (300 seconds), and each nominally involves complete cycles from the high setting of the fan down to its minimum setting over the course of 11 setting points that were determined to be roughly comparable in a preliminary effort, and then back up to the original maximum setting. Records were retained that were free of spurious outliers (usually corresponding to bumping the apparatus). Three 300-second time records were taken, and exponential functions were fit to the relationship between the smoothed velocity and vibration power time series one as shown in Fig.~\ref{fig:fits}.

Using the fits derived from each one, we can compare the predictions produced by each fit with the data not used to produce the fit, and in doing so determine the expected predictive accuracy via cross validation; these results are shown in Table~\ref{tab:cross_validation}. The diagonal terms are the cases where the same data was used to fit and evaluate the model, and so should be disregarded for cross validation purposes, but are shown for the sake of completeness. The largest RMS prediction error is only 0.1818~m/s, and the mean is 0.1544~m/s, which seems rather modest considering that the experiment was conducted in an electrically and acoustically noisy environment, used a single sensor, and was fundamentally a preliminary effort. Given these considerations, it seems likely that further refinement of this approach with an ensemble of sensors and a generally quieter environment could yield predictions with accuracies an order of magnitude better.
\begin{table}[]
    \centering
    \begin{tabular}{c|c|c|c}
    &set 1&set 2& set 3\\
    \hline
    set 1&\textcolor{red}{0.0870}&0.1149&0.1750\\
    set 2&0.1202&\textcolor{red}{0.1018}&0.1786\\
    set 3&0.1562&0.1818&\textcolor{red}{0.0712}\\
    \end{tabular}
    \caption{Cross validation predictive accuracy in RMS error for three 300-second data sets each containing a single cycle of the fan settings.}
    \label{tab:cross_validation}
\end{table}

\section{Concluding discussion}
For the water case it was possible to resolve linear velocity differences on the order of a cm/second. For the air case the mean RMS error was 15.4~cm/s both of which are fairly narrow velocity uncertainties, and accurate enough to be useful. In the case of the water measurement, the vibration-based measure was able to consistently resolve a difference in velocity with the highest two pump setting smaller than could be detected using the gauge we employed to measure the velocity. In both cases, the application of this approach to flows outside rather than inside a cylinder can open up further application areas, but requires further development to generalize this approach to these cases. It is important to note that the method we here employ is expected to be applicable to fluids with constant parameters; e.g., density, viscosity, etc, and mixtures, or fluids with changing parameters might lead to spurious results. For the navigational application specifically, merging data from both vibration sensors and an INS through Kalman filtering could be expected to achieve superior combined performance during periods of change, providing a path to successful transient performance in additional to the successful resolution of flow velocities demonstrated here for more steady cases. Additionally, in this study we have only used a single sensor on each tube; additional approaches are possible using a line array of sensors that could potentially increase both accuracy and ability to distinguish between velocity changes and other parametric changes in the consistency of the fluid under consideration.

\section{Acknowledgments}
Sandia National Laboratories is a multi-mission laboratory managed and operated by National Technology \& Engineering Solutions of Sandia, LLC (NTESS), a wholly owned subsidiary of Honeywell International Inc., for the U.S. Department of Energy's National Nuclear Security Administration (DOE/NNSA) under contract DE-NA0003525. This written work is authored by an employee of NTESS. The employee, not NTESS, owns the right, title, and interest in and to the written work and is responsible for its contents. Any subjective views or opinions that might be expressed in the written work do not necessarily represent the views of the U.S. Government. The publisher acknowledges that the U.S. Government retains a non-exclusive, paid-up, irrevocable, world-wide license to publish or reproduce the published form of this written work or allow others to do so for U.S. Government purposes. The DOE will provide public access to results of federally sponsored research in accordance with the DOE Public Access Plan.

\bibliography{aipsamp}

@PREAMBLE{
 "\providecommand{\noopsort}[1]{}" 
 # "\providecommand{\singleletter}[1]{#1}%" 
}

@ARTICLE{Ren:2022,
   author       = "R.~Ren and H.~Wang and X.~Sun and H.~Quan",
   year         = "2022",
   journal      = "Sensors",
   volume       = "22",
   pages        = "7470",
}

@ARTICLE{Korlapati:2022,
   author       = "N.~V.~S.~Korlapati and F.~Khan and Q.~Noor and S.~Mirza and S.~Vaddiraju",
   year         = "2022",
   journal      = "J. of Pipeline Sci. Eng.",
   volume       = "2",
   pages        = "100074",
}

@ARTICLE{Smith:2026,
   author       = "C.~B.~Smith and S.~H.~Swift and A.~Steyer and I.~El-Kady",
   year         = "2026",
   journal      = "arXiv",
   volume       = "",
   pages        = "arXiv:2603.23496",
}

@article{Swift:2025,
  author={S.~H.~Swift and I.~F.~El-Kady},
  journal={J. Acoust. Soc. Am.},
  volume={157},
  number={4\_Supplement},
  pages={A280--A280},
  year={2025},
  publisher={Acoustical Society of America}
}

@ARTICLE{Evans:2004,
   author       = "R.~P.~Evans and J.~D.~Blotter and A.~G.~Stephens",
   year         = "2004",
   journal      = "J. Fluids Eng.",
   volume       = "126",
   pages        = "280-285",
}

@ARTICLE{Kim:1996,
   author       = "Y.-K.~Kim and Y.-H.~Kim",
   year         = "1996",
   journal      = "J. Acoust. Soc. Am.",
   volume       = "100",
   pages        = "717-726",
}

@TECHREPORT{Brody:2023,
   author = "J.~Brody and S.~Leung and J.~Shamwell and D.~Donavanik",
   title = "Sources of Error in GPS-Denied Odometry",
   institution = "DEVCOM Army Research Laboratory",
   type = "Report",
   number = "ARL-TR-9798",
   address = "DEVCOM, Army Research Laboratory, Aberdeen Proving Ground, MD",
   month = Sept,
   year = 2023,
   note = "A full TECHREPORT entry",
}

\end{document}